\documentclass[sigconf]{acmart}

\usepackage{booktabs} 
\usepackage{multirow}
\usepackage{times}
\usepackage{epsfig}
\usepackage{graphicx}
\usepackage{amsmath}
\usepackage{amssymb}
\usepackage{caption}
\usepackage{subfig}
\usepackage{tabularx}
\usepackage{algorithm}
\usepackage[noend]{algpseudocode}
\usepackage{bchart}
\usepackage{pgfplots}
\usepackage[font=small]{caption}
\usepackage[T1]{fontenc}
\usepackage{eso-pic}
\usepackage{pbox}
\usepackage{ragged2e}
\usepackage{xspace}

\hypersetup{
  colorlinks = true,
  citecolor = green
}

\newcommand{\footlabel}[2]{%
    \addtocounter{footnote}{1}%
    \footnotetext[\thefootnote]{%
        \addtocounter{footnote}{-1}%
        \refstepcounter{footnote}\label{#1}%
        #2%
    }%
    $^{\ref{#1}}$%
}

\newcommand{\footref}[1]{%
    $^{\ref{#1}}$%
}

\makeatletter
\renewcommand\@makefnmark{\hbox{\@textsuperscript{\normalfont\color{red}\@thefnmark}}}
\makeatother

\AddToShipoutPicture{%
  \AtTextUpperLeft{
    \put(0,\LenToUnit{1.5cm}){\parbox{\textwidth}{\centering
       This paper was accepted to the ACM's Special Interest Group on Information Retrieval (SIGIR) workshops, 2018.}}
  }
}

\setcopyright{rightsretained}
\acmConference[SIGIR 2018 eCom]{ACM SIGIR Workshop on eCommerce}{July 2018}{Ann Arbor, Michigan, USA} 
\acmYear{2018}
\copyrightyear{2018}

\begin{document}
\title{A Multimodal Recommender System for Large-scale Assortment Generation in E-commerce}
\renewcommand{\shorttitle}{Assortment Recommendations}


 \author{Murium Iqbal}
 \affiliation{%
   \institution{Overstock}
   \city{Midvale}
   \state{Utah}
 }
 \email{miqbal@Overstock.com}

 \author{Adair Kovac}
 \affiliation{%
   \institution{Overstock}
   \city{Midvale}
   \state{Utah}
 }
 \email{akovac@Overstock.com}

\author{Kamelia Aryafar}
 \affiliation{%
   \institution{Overstock}
   \city{Midvale}
   \state{Utah}
 }
 \email{karyafar@Overstock.com}


\begin{abstract}
E-commerce platforms surface interesting products largely through product recommendations that capture users' styles and aesthetic preferences. Curating recommendations as a complete complementary set, or assortment, is critical for a successful e-commerce experience, especially for product categories such as furniture, where items are selected together with the overall theme, style or ambiance of a space in mind. In this paper, we propose two visually-aware recommender systems that can automatically curate an assortment of living room furniture around a couple of pre-selected seed pieces for the room. The first system aims to maximize the visual-based style compatibility of the entire selection by making use of transfer learning and topic modeling. The second system extends the first by incorporating text data and applying polylingual topic modeling to infer style over both modalities. We review the production pipeline for surfacing these visually-aware recommender systems and compare them through offline validations and large-scale online A/B tests on Overstock~\footlabel{overstockurl}{\url{www.overstock.com}}. Our experimental results show that complimentary style is best discovered over product sets when both visual and textual data are incorporated.
\end{abstract}

%
%



\maketitle

\section{Introduction}
\label{sec:intro}

Overstock~\footref{overstockurl} is an e-commerce platform with the goal of creating \textit{dream homes for all}. Users browse Overstock's catalog to select pieces that complement one another while matching the stylistic settings and color palettes of their rooms. As furniture is not a disposable product, furniture purchases are subject to careful scrutiny of aesthetics and strict budgeting. Brick and mortar stores often inspire consumers by creating furniture showrooms, in some cases pushing consumers to walk through their carefully selected furniture displays. These assorted showrooms alleviate the creative and stylistic pressure on the consumer, since the set is already on display with all the necessary pieces. 

\begin{figure}[t!]
\centering
        \includegraphics[scale=0.50]{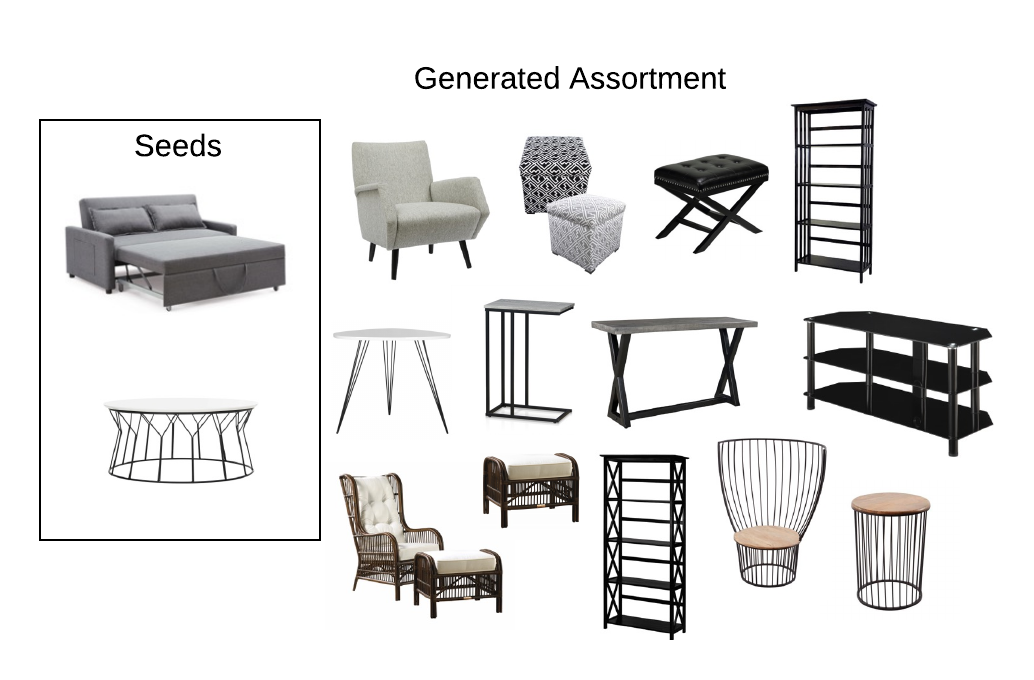}

\caption{An automatically generated assortment from the multimodal approach is shown.}
\label{fig:assortmentJaccard}
\end{figure}

By crafting an appropriate recommender system, this experience can be recreated on an e-commerce platform. Figure~\ref{fig:assortmentJaccard} illustrates an example of the proposed system's automatically generated showroom, built around two seed products selected from Overstock~\footref{overstockurl}. The goal of this recommender system is to provide set recommendations that adhere to a general theme or cohesive visual style while accounting for essential item constraints~\footnote{Essential item constraints refers to specific products which must be included in an assortment, \textit{e.g.} a bed frame in a bedroom set or a vendor must-have in a subscription box.}. Set recommendations have become more prominent with the rise of subscription box services in various domains such as fashion (\textit{e.g.}~StitchFix~\footnote{\url{www.stitchfix.com}}), jewelry (\textit{e.g.}~Rocksbox~\footnote{\url{www.rocksbox.com}}) and beauty products (\textit{e.g.}~Birchbox~\footnote{\url{www.birchbox.com}}). The criteria for selecting an assortment depends heavily on the product space which can inform the latent space of user preferences and product representations. Personal care product recommendations, for example, rely on user preferences, such as skin type, which can be inferred from product text-based attributes. Fashion, jewelry and furniture shopping, on the other hand, are predominantly visual experiences. This motivates a \textit{visually-aware representation} for products and style-based preferences for users. 

We propose two assortment recommender systems. The first takes advantage of visually semantic features transferred from a deep convolutional neural network to learn style. The second uses both these visual features and product text-based attributes to learn style across the multimodal dataset. Our hypothesis is that while the visual style can help find similar products, use of text data in conjunction with images will result in more complimentary, and cohesive stylized assortments. 

\begin{figure}[t!]
\centering
        \includegraphics[scale=0.70]{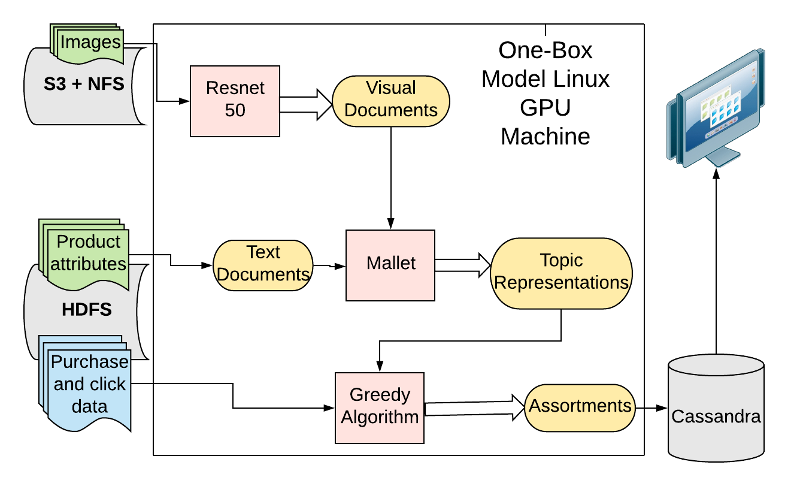}

\caption{A system overview diagram is presented above. We use product data and user engagement data residing within our Hadoop cluster and our NFS/S3 store to train the model. Most of the training is done on a one box GPU server, which runs the {\tt Resnet-50}, Mallet and our greedy algorithm. All results are served to Overstock~\footref{overstockurl} users from Cassandra.}
\label{fig:Systemoverview}
\end{figure}

\section{Related Work}
\label{sec:related_work}

As the prevalence of recommendation systems grows, e-commerce platforms are looking to increase the influence on customer purchases by not just recommending one item at a time but several in a bundle together \cite{xie2010breaking} \cite{zhu2014bundle} \cite{beladev2016recommender} \cite{garfinkel2006design}. This is true across industries, from travel agencies to clothing \cite{hsiao2017creating} \cite{liu2011personalized} \cite{liu2012hi}. Specifically Zhu et al.\cite{zhu2014bundle} model the problem of bundle recommendations as the Quadratic Knapsack Problem (QKP), and find an approximation to it. They use only implicit feedback data, rather than content data to build their system, allowing for the bundles to be personalized. Using this approach, we also view our system as a QKP, but look to leverage primarily content data, incorporating implicit feedback as a minor signal. To capture a user's preference, the system we propose allows a user to present a seed product, around which we can create the bundle. This allows us to circumvent the cold start problem, as any user, regardless of having a history on our site, can still build assortment recommendations. New products are also seamlessly incorporated, as long as images or text data is made available. 

Recommendation systems is a well studied field which has two large over arching types: collaborative filtering methods \cite{mnih2008probabilistic} \cite{hu2008collaborative} \cite{gopalan2015scalable}, and content based methods \cite{cunha2017metalearning} \cite{van2013deep} \cite{van2000using} of which many rely on learning latent representation of the input data. Topic modeling, specifically Latent Dirichlet Allocation (LDA)\cite{blei2003latent}, has also been applied to create topic based recommendations both off of text input as well as from implicit feedback \cite{hu2014style} \cite{bergamaschi2014comparing} \cite{xie2015recommendation}. These systems work by scoring individual products against one another or against users to see which are most similar. Here we look to extend this methodology by applying topic modeling over both image and text data. Rather than simply concatenating image and text features and linearly combining them, PolyLDA enables us to learn two distinct but coupled latent style representations. This allows for a versatile interpretation of style \cite{iqbal2018discovering}. We then use these learned styles to make bundle or assortment recommendations rather than the traditional single item recommendations. 

To facilitate style discovery from the images, we process them via prevalent deep learning techniques. Deep residual networks have recently been shown as a powerful model to capture style based preferences for creating visual wardrobes from fashion images~\cite{hsiao2017creating}. With the goal of learning visually coherent styles, we apply transfer learning, by using a {\tt Resnet-50}~\footnote{\url{https://github.com/facebook/fb.resnet.torch}} which was pre-trained on ImageNet~\footnote{\url{http://www.image-net.org/}}~\cite{he2016deep}. We explore interpreting the convolutional neural network by indexing the activations on channels within the convolutional layers~\cite{rafegas2017understanding}. We use a neural network to learn filters for our data and simply index their responses to images to create visual documents, similar to the older bag of word methods~\cite{weinberger2009feature}. To discover visually-aware trends, we use these documents with Latent Dirichlet Allocation (LDA)~\cite{blei2003latent}. 

While we use LDA for topic modeling on single-modality image data, we need an extension to interpret both visual semantic features and text-based attribute data as complementary modalities. Roller et al. ~\cite{roller2013multimodal} create a multimodal topic model that assumes words and corresponding visual features occur in pairs and should be captured as tuples within the topic distribution. This model would work well for descriptions of images, but the underlying assumption could prove to be too stringent to generalize to our application. Mimno et al.~\cite{mimno2009polylingual} offer a more flexible extension of LDA, Polylingual LDA (PolyLDA), which handles tuples of documents in different languages with assumed identical topic distributions. The documents within a tuple do not need to be direct translations of one another, and the topics themselves have distinct sets of words for each language. By treating our image data and our text data as two separate languages, we can use PolyLDA directly to create a more flexible multimodal topic model with which to infer style.

Hsiao and Grauman \cite{hsiao2017creating} also apply PolyLDA to infer style from images, but they use complementary clothing types (e.g. pants and blouse) as languages in a compatibility model instead of using PolyLDA to handle different modalities of data for the same item. The documents they use for their topic models are generated using a Resnet trained to automatically detect their pre-defined text attributes within images and apply the appropriate labels.

Much of the work performed on traditional recommendation systems can be applied here, but additional constraints must be taken into consideration. Products can't simply be visually similar to be purchased together, but must be complimentary as well. For example, having two similar brightly colored upholstered chairs as a result in a traditional recommendation is perfectly acceptable, but if they are both taken into one assortment with the intention of being bought together, they may clash. Here we show that by incorporating text data, we are able to avoid assortments of incredibly visually similar, interchangeable products which are not at all complimentary. These nuances can be further addressed by incorporating purchase data to infer what types of products are not just similar, but also compatible, and likely to be bought together. To create the final assortment recommendations, we define a distance measure that combines stylistic similarity with implicit feedback from users, in the form of purchases, to develop a concept of affinity similar to approaches in composite recommendation systems such as ~\cite{zhu2014bundle,hsiao2017creating,veit2015learning,he2016learning}.

This rest of this paper is organized as follows: In Section~\ref{sec:method}, we present our production pipeline for assortment recommendations as deployed in Overstock~\footref{overstockurl} and explore our product embeddings as used with topic modeling techniques. We also review our formulation of assortment generation as a 0-1 quadratic knapsack problem and our greedy optimization for building assortments. Section~\ref{sec:results} presents our offline and online experimental results. Finally, we conclude our paper in Section~\ref{sec:conclusion} and propose future directions.


\section{Methodology}
\label{sec:method}

Here we provide an overview of the production system used to surface recommendations throughout Overstock~\footref{overstockurl} and describe our assortment recommender system in detail. We first 
explain our product embeddings as text-based bag-of-words and bag-of-visual-words (BoVW) documents. Then we explain the LDA approaches used to discover text-based and visual styles across our platform. We finally discuss our formulation of assortment generation as a 0-1 quadratic knapsack problem with budget constraints and our greedy assortment recommendations algorithm. 

\subsection{System Overview}\label{sec:system_overview}

\begin{figure}[t!]
\centering
        \includegraphics[scale=0.5]{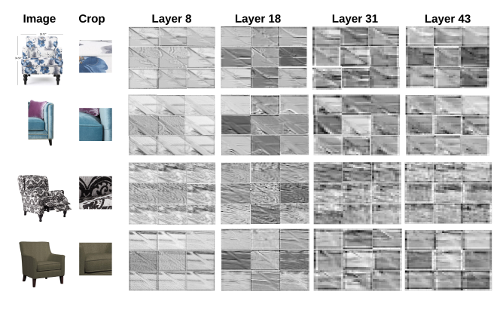}

\caption{Here we depict activations from the layers used to create our visual documents. The leftmost column is the raw image. The next column to the right is the cropped patch which is passed through the network. In the next 4 columns, we see the response to these patches from 9 channels each of convolutional layers 8, 18, 31 and 43. Lighter colors mean strong responses, while darker colors mean weak responses. We threshold these responses to determine whether or not the "word" each channel represents is present within the image. This methodology seems to work particularly well for our upholstered furniture, showing distinct responses to the different patterns.}
\label{fig:ResnetActivations}
\end{figure}

Product recommendations on Overstock~\footref{overstockurl} are surfaced through multiple carousels across the platform.  User interactions with products and recommendations is recorded along side purchase data. Each product on the platform is represented by corresponding title, text attribute tags, text-based descriptions and images. The first step in generating recommendations is processing the large volume of product and user data which are hosted on Hadoop Servers.

After initial processing on Hadoop, the data is transfered to a one-box CUDA-enabled GPU server. Here product images are passed through a publicly available {\tt Resnet-50} to create image-based documents. Both the image and text documents are then fed into Mallet~\footnote{\url{http://mallet.cs.umass.edu/}} to generate topic representations of products. We create two topic variants, a visual variant, that uses the image data with LDA and a multimodal variant, that uses both text and image data with PolyLDA. 

Once the topic distributions are created, the products reside within their defined feature space. Hereafter, the methodology for both variants is identical. We construct a distance measure over the topic space, using implicit feedback from our users, to model compatibility. The assortments are then generated by finding products which minimize this distance to predefined seed product pairs.

The generated assortments are pushed to our Cassandra database cluster. Our website servers can query Cassandra directly to see which products have an associated assortment to present to users. Assortment recommendations are featured on corresponding product pages, below the product description, providing an option for users to purchase an entire assortment, or just parts of it. A brief diagram of the systems involved is presented in Figure ~\ref{fig:Systemoverview}.

\subsection{Product Embeddings}\label{sec:document_creation}

Products on Overstock~\footref{overstockurl} have several images of the item for sale and different forms of associated text data including title, description and attribute tags. Product attributes are descriptive tags associated with a product. Some example attribute categories are color, size, style, composing materials, finish, and brand. In this paper, we utilize these attributes as the primary text-based information since they often provide a rich text representation of the item that compliments provided images. All product attribute data is processed to remove stop words and are concatenated with each product title to form a bag-of-words document.

We use the methodology described in \cite{iqbal2018discovering} to create visual documents to be used with LDA via a Resnet. We then combine both visual and text data together via PolyLDA.

The convolutional layers within a Resnet are composed of a series of small learned filters. Each filter is convolved with the input in steps along it's height and width. The result of this process is a 2-dimensional grid of activations that represents the response from the filter at corresponding locations of the input. The learned filters within a Resnet respond to specific patterns. By viewing these filters as words, and indexing them, Iqbal et al. \cite{iqbal2018discovering} show that a BoVW document can be created which, when combined with LDA, is effective at uncovering style. To index a layer, we simply threshold the response from the filter to indicate whether or not it is sufficiently activated by the image to be included within the image's document.

\begin{figure}[t!]
\centering
        \includegraphics[scale=0.80]{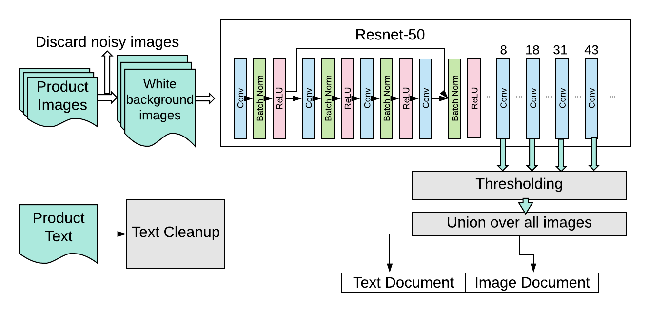}

\caption{Each product's corresponding text and images are collected. Our system then discards images with noisy scenes, keeping only those with white backgrounds and passing them through {\tt Resnet-50}. Channel activations from layers 8,18,31, and 43 are thresholded and indexed to create visual documents. The union of these documents is taken as the visual representation of the product. The text data is stripped of stopwords and paired with the corresponding visual document to create a tuple for ingestion by PolyLDA}
\label{fig:Documentcreation}
\end{figure}

Given this methodology for creating image documents, our visual word vocabulary is defined entirely by the channels in the layers that we choose to index. We tested creating documents using various layers through the network. We found that combining several middle layers gave us the best results. After experimenting we empirically chose to use {\tt Resnet-50} layers 8, 18, 31, and 43. The combination of these layers provided significantly better results, per visual inspection, than any single layer or other combinations. This yielded a total vocabulary size of 2816 visual words over our image documents. Since multiple images can be provided for each  product, we take the union of all present visual words within the corresponding images as the visual document for the product itself. The process is depicted in Figure \ref{fig:Documentcreation}.

\subsection{Topic Modeling}\label{sec:topic_modeling}
Once the visual documents are created, they can be used directly with LDA to create topics for our visual variant. LDA is an unsupervised generative model which infers probability distributions of words within topics and models documents as a mixture of those topics \cite{blei2003latent}. Given a set of documents and a number of target topics the model assumes the following generative process was used to create the documents.

\begin{enumerate}
\item For $M$ documents, initialize the set of topic distributions, $\theta_i \sim \mathbf{Dir(\alpha)}$ for \{$i=1\dots $M$\}$
\item For $K$ topics, initialize the set of word distributions, ${\phi_k} \sim \mathbf{Dir(\beta)}$  for \{$k=1\dots $K$\}$ 
\item For the $k^{th}$ word in the $i^{th}$ document select a topic ${z_i}$ from  $\theta_i$ and a word, ${w_{i,k}}$, from ${\phi_{z_i}}$
\end{enumerate}
where $\mathbf{Dir}$ is the Dirichlet distribution.

The model initializes the parameters randomly then iteratively updates these parameters via Gibbs Sampling and variational Bayes inference. After many iterations the topic distributions converge to a stable state and the resulting topics can be used as a low dimensional feature space that captures the salient content of the original documents.

For the multimodal variant, each visual document is paired with it's text equivalent in a tuple. These tuples are then used with Polylingual LDA (PolyLDA) \cite{mimno2009polylingual}. Products that are missing either a text document or a visual document can still be used with PolyLDA, as it is robust to missing documents within tuples. A diagram of the system for multimodal document creation is provided in Figure~\ref{fig:Documentcreation}. PolyLDA is an extension of LDA meant to handle loosely equivalent documents in different languages. It takes as input tuples of documents, each in a different language but with the same exact topic distribution. Each resulting latent topic from this method contains a distinct associated set of word distributions for each language. Documents aren't required to be direct translations of one another, which allows for flexibility. A word in one language could be attributed to a given topic while its direct translation in another language may be attributed to a different topic. Words that only appear in one language can also be present in topics despite having no equivalent translations in the corpus. PolyLDA extends LDA by assuming the following generative procedure:

\begin{enumerate}
\item For a given tuple of documents, $\{d_1\dots d_L\}$, initialize a single set of topic distributions, $\theta \sim \mathbf{Dir(\alpha)}$
\item For $K$ topic sets with $L$ languages, initialize the set of word distributions, ${\phi_{k,l}} \sim \mathbf{Dir(\beta)}$  for $\{k=1\dots K\}$ and $\{l=1\dots L\}$ 
\item For the $k^{th}$ word in the $l^{th}$ document in the $i^{th}$ tuple select a topic ${z_{i,l}}$ from  $\theta_i$ and a word, ${w_{i,l,k}}$, from ${\phi_{z_{i,l}}}$
\end{enumerate}

For our multimodal variant, we make use of the flexibility afforded by PolyLDA by applying it to our visual and text documents, representing each modality as a distinct language attempting to describe the same product. This method should allow for topics that afford better generality. Intuitively for our corpus, we suspect that a topic could capture relationships between certain visual features, like thick wood, with textual attributes, like "quality", that don't have a direct visual representation but do share an underlying contextual relationship. By combining modalities the model also affords us the ability to infer a more complete style representation given incomplete data. For example, we can apply our multimodal topics to products with only text attributes available and still infer style that would normally be captured by the missing image data. 

\begin{figure*}[t!]
        \centering
\subfloat{%
  \includegraphics[scale=0.70] {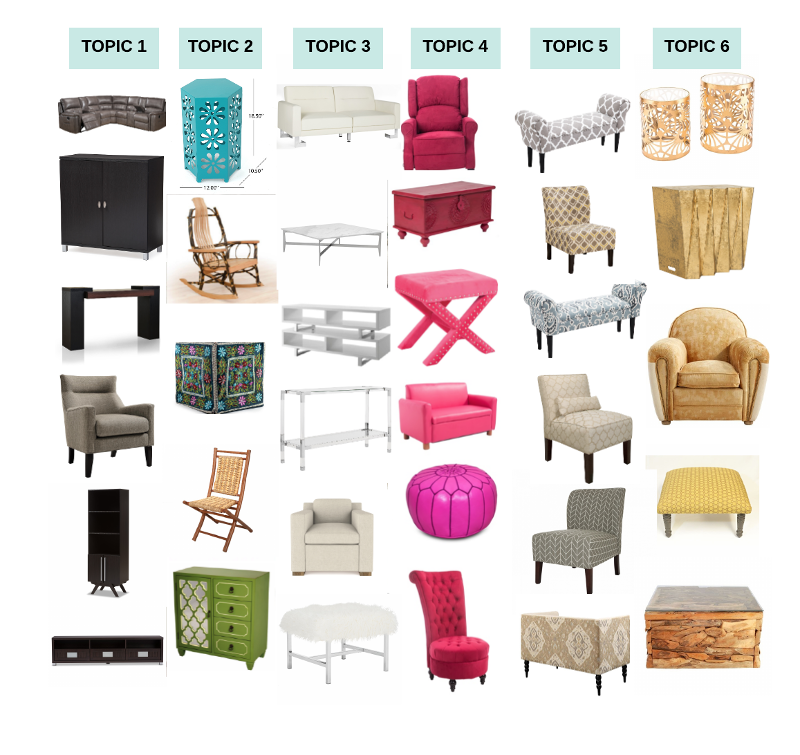}\hfill
}
\subfloat{%
  \includegraphics[scale=0.75] {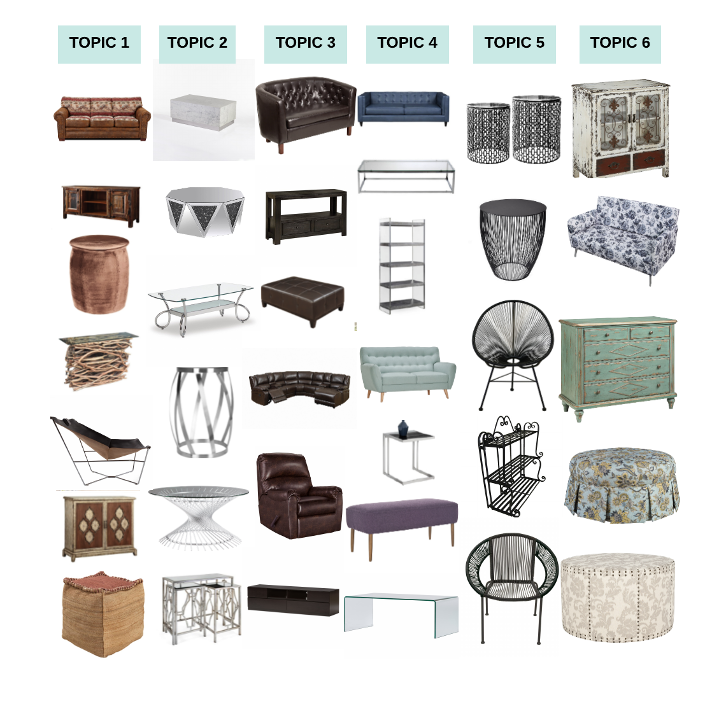}
}
         \centering
        \caption{Representative items from each of six topics in the visual variant on the left (a) and multimodal variant on the right (b) are depicted above.}
        \label{figure:topics}

\end{figure*}

We provide visualizations of the resulting topics from both variants in Figure ~\ref{figure:topics}. To create these visualizations we selected the products that are maximally aligned with each topic. In our case, we have empirically seen that the topics relate well to various styles while reducing our feature space from a large number of attribute tags, title words, and image-based channel activations to a succinct $k$-dimensional space (where $k$ is the number of topics). These styles are not necessarily complimentary though, especially in the case of the visual topics. We can see that the items within the topics seem more substitutable, e.g. in the cases of brightly colored furniture, similarly patterned upholstery, and mirrored surfaces.

\subsection{Knapsack Problem}\label{sec:greedy_method}

Once we have all our products residing within the topics' feature space, we can begin to build out our assortments. The process for generating the assortments is the same for both the visual variant and the multimodal variant, with the only difference between the two being the topic representations of products. To build these assortments, we first define the necessary components of an assortment as verticals. The verticals we define here enable us to build living room assortments, but we would like to emphasize that verticals can be defined for any e-commerce platform. If desired, the verticals can be left undefined, and the assortment can be built with no vertical constraints. For our specific application verticals are critical pieces of living room furniture, e.g. coffee table, chair, or accent table. Our definitions are manually applied to the products and are listed below. (Here a couch set can be either a sectional sofa or a sofa and loveseat combination.)

\paragraph{Verticals:}
\{Couch Set, Coffee Table, Accent Table, Entertainment Center, Bookshelf, Ottoman, Chair\}
\\

We chose to build our assortments around seed pieces. For our current application, we generated seeds, but we can easily allow users to provide their own seeds around which we can curate an assortment. We selected seeds as preferred pairings of the most crucial verticals in any living room space, namely a couch set and a coffee table. Although all the other verticals can be seen as optional, these two verticals are defining members of a living room and therefore must be present in any assortment. We chose pairs of sofas and coffee tables which are most frequently co-clicked for our seeds.

We then assume a budget constraint for the entire assortment. As such we can formulate our assortment generation as a 0-1 quadratic knapsack problem (0-1 QKP) as defined by Gallo et al \cite{gallo1980quadratic}. The 0-1 QKP assumes that a knapsack with limited space needs to be filled with a set of items. These items all have an intrinsic value. Items also have an additional associated pair-wise profit when selected together for the knapsack. 

For a given seed, we can find the optimal assortments by trying to find the products that are nearest the seed in the topic space. We can add the additional constraint that we want all products within the assortment to be nearest one another in the topic space as well. The total assortment must remain within the assumed budget constraint. Each product has an associated vertical, and we can add in some constraints on the minimum and maximum number of products each vertical can contribute to the final assortment.

To calculate proximity we use a Mahalanobis distance built on purchase data similar to \cite{mcauley2015image}. The Mahalanobis distance allows us to leverage the implicit feedback of our users to understand how our topics distribute and relate to one another across multiple products within a purchase. 

We assume that purchases of living room furniture made by the same user over a 3 month period are being used in the same room, and as such can be said to represent a compatible assortment. After trimming to include only purchases with 3-10 pieces of living room furniture, so as to avoid noise and erroneous signals from bulk purchasers, we create a dataset of roughly 7.5k purchased assortments. We take the $L_2$ normalized sum of the topic distributions of the co-purchased products as a topic representation of the purchase itself, and use this dataset to calculate the covariance matrix used for our Mahalanobis distance. 

The Mahalanobis distance is defined as $d_M$($x_i$,$x_j$) = ($x_i$-$x_j$)$\mathbf{M}$($x_i$-$x_j$)$^T$, Where $\mathbf{M}$ is the covariance matrix for the topics learned from our purchase data. For each seed pair of coffee table and couch set, we greedily swap products in and out of our assortment until we converge to an ideal set of products. The only verticals which are not allowed to change are those in the seed.

\begin{algorithm}[t!]
\caption{Generate Assortments}
\begin{algorithmic}[2]
\State $\delta \gets \epsilon + 1$
\State ${A_{i,prev}} \gets \{\} \forall ${i}$ \in\{$V$ - $S$\} $
\State ${A_{i,prev}} \gets $\{${p_{i}}\} \forall ${i}$ \in\{$S$\} $
\While{$\delta \geq \epsilon $}
\State $\delta \gets 0$
\For {${i} \in\{$V$ - $S$\} $}
\State ${t_A} \gets \Vert (\sum t_p) \Vert$ $\forall {p_j} \in {A_{j,prev}}$, $\forall j \in {V} \neq {i}$
\While{$ \mathbf{size}({A_i}) < {i_{size}} $}
\State ${A_i} \gets {A_i} \cup \mathbf{argmin} (\mathbf{d_M}({t_o},{t_A}))$
\EndWhile
\State $\delta \gets \delta + {d_M}({t_{A_i}},{t_{A_{i,prev}}})$
\EndFor
\EndWhile
\end{algorithmic}
\label{alg:generateAssortments}
\end{algorithm}

To formulate the O-1 QKP we first define some terminology. Each product $o$ will have a vertical label $a_o$, an associated price $c_o$, and associated topic distribution $t_o$. All products will have a pairwise distance associated with them and every other product ${d_{i,j}}$ which is the Mahalanobis distance between products $o_i$ and $o_j$ built on the purchases and topic distributions. Additionally the products all have a score $q_o$ to represent their value to the seed, here we use the inverse of the distance from the product to the seed. Let $M$ be the set of products comprising the assortment. Let $B$ be the total available budget minus the cost of the seed. The 0-1 knapsack problem can then be formulated as:

\begin{eqnarray}
\begin{aligned}
\label{QKP}
\mbox{ maximize  }  \sum_{i=1}^M {q_{i}} + \sum_{j=1}^M 1 / {d_{i,j}}, j \neq i\\
\text{ subject to:} \\
\forall~i:~\sum_{i=1}^M {c_i} \leq  B \\
\forall~k:\,\,{min_k}\leq \mathbf{count}({a_o} = k) \leq {max_k}
\end{aligned}
\end{eqnarray}

A greedy approximation can now be formulated to follow the constraints of the 0-1 QKP. First a solution is initialized, then iteratively, items are swapped until the system converges to an optimal solution. To get the initial set, products are all scored by their total potential for compatibility (the sum of the score of this product with the sum total of all reciprocals of pairwise Mahalanobis distance) divided by their price. Products are then sorted by this ratio and the highest scored are added until vertical constraints or budget constraints are met. Then each product is considered for a swap with other products to improve the total compatibility of the assortment.

\subsection{Greedy Method}\label{sec:greedy_method}

\begin{figure*}[t!]
        \centering
\subfloat[][]{%
                \fbox{\includegraphics[height=4cm] {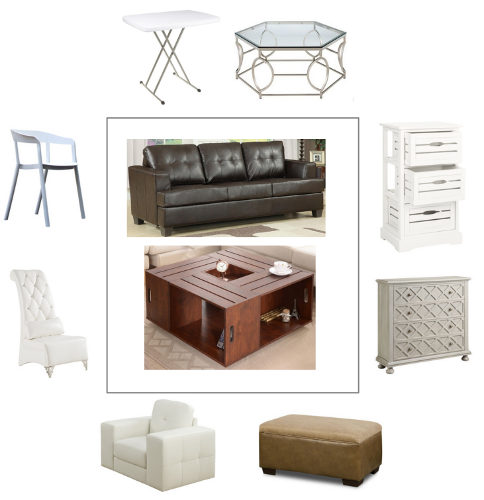}}
       }
\subfloat[][]{%
                \fbox{\includegraphics[height=4cm] {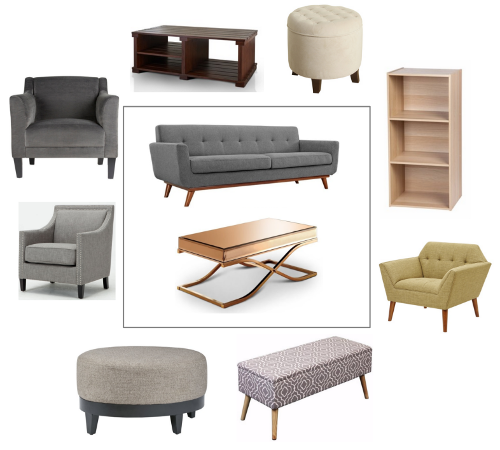}}
       }
\subfloat[][]{%
                \fbox{\includegraphics[height=4cm] {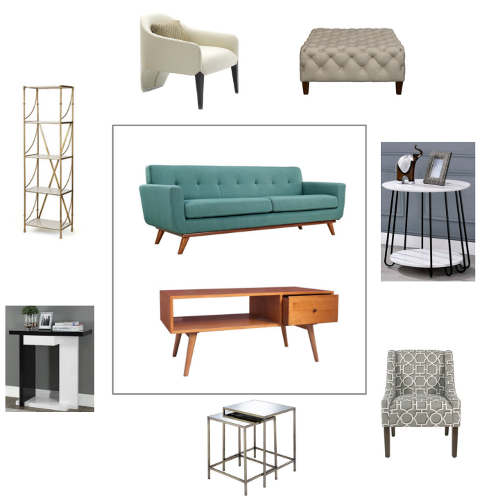}}
       }
\subfloat[][]{%
                \fbox{\includegraphics[height=4cm] {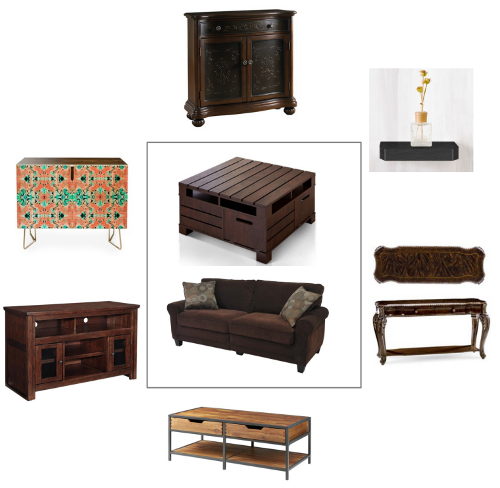}}
       }\\
\subfloat[][]{%
                \fbox{\includegraphics[height=3.65cm] {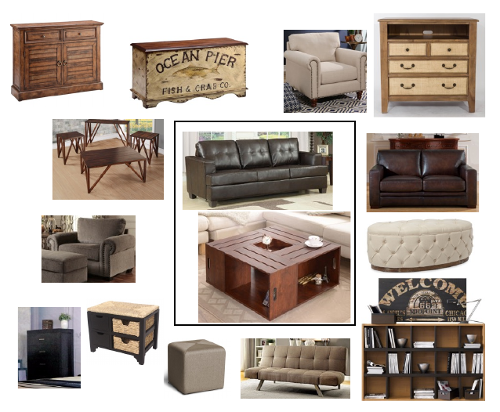}}
       }
\subfloat[][]{%
                \fbox{\includegraphics[height=3.65cm] {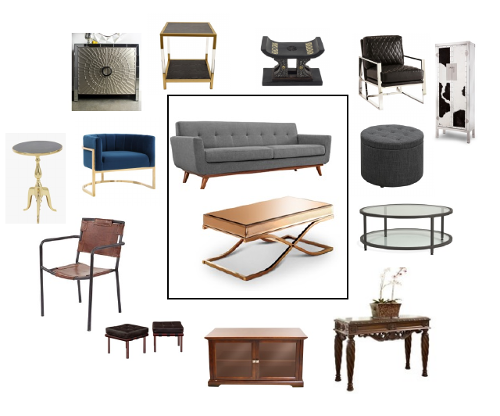}}
       }       
\subfloat[][]{%
                \fbox{\includegraphics[height=3.65cm] {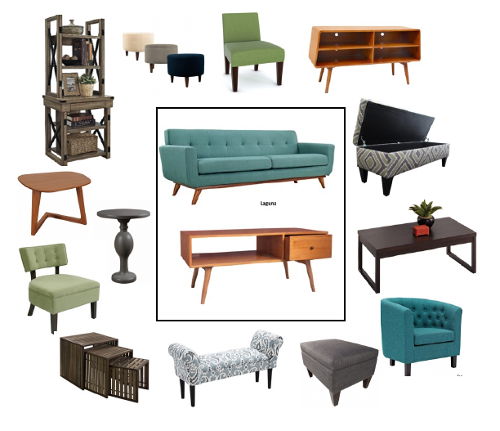}}
       }       
\subfloat[][]{%
                \fbox{\includegraphics[height=3.65cm] {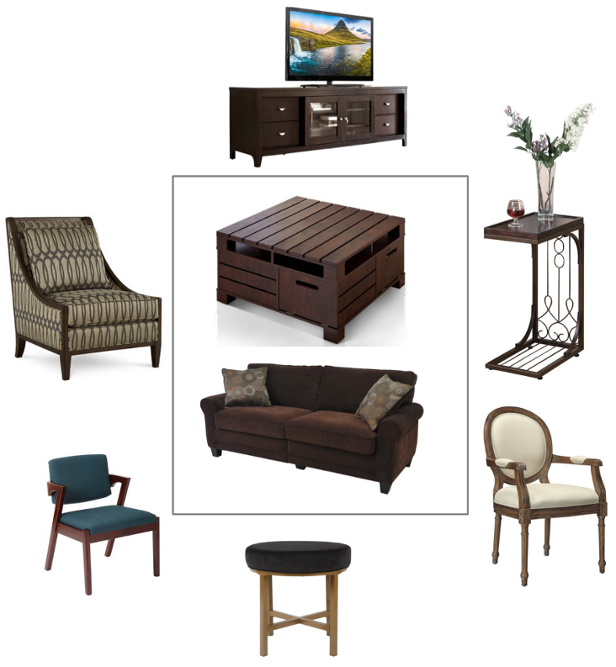}}
       }
         \centering
        \caption{Assortments from our visual variant (top) and multimodal variant (bottom) with the same seeds are depicted above.  The visual assortments tend to provide a more-of-the-same assortment, while the multimodal is able to create a diverse set of products that still forms a cohesive style. When the visual assortment can't find products similar to either seed, it selects products which are similar to one another. In some cases it can match products to one seed, and ignores the other, which leaves the second seed item looking out of place. The multimodal variant, which can take advantage of text data as well, is able to find pieces to create a cohesive look incorporating elements of both seeds.}
        \label{figure:collections}

\end{figure*}

\newcommand{\tuple}[1]{\ensuremath{\left \langle #1 \right \rangle }}

For our online evaluation, we relax the budget constraints, so that we can display the assortment on product pages. This allows us to select the best assortment per product. We can easily reapply the budget constraint for user assortments. As such, we use the following greedy method to generate the assortments. The assortment is initialized as first only containing the seed items, all other verticals are empty. Each vertical is then iteratively considered for new products while holding all other verticals constant. New products are chosen from the pool of products that are labeled as candidates for this vertical. Those which are closest to the total assortment (minus the current vertical) at the previous time-step are added in, until the size of the vertical is met. The sizes of the verticals in our experiment were set by us, but a user can dictate how many side tables, bookshelves, chairs, etc they wish to add to their assortment.

Let $V$ be the set of all verticals. $S$ is the set of verticals used in the seed. $A_i$ is the set of products selected for the assortment in vertical $i$ for the current timestep. Each vertical $i$ has an associated size, ${i_{size}}$, which is the number of products we want for this vertical. $A_{i,prev}$ is the set of products selected for the assortment in vertical $i$ for the previous timestep. $p_i$ is an element of $A_{i,prev}$. $o_i$ is a candidate product for vertical $i$. $t_o$ is the topic distribution of a product. $d_M$ is the Mahalanobis distance as defined above. $\delta$ is the change in the distance of the assortment from iteration to iteration. The greedy approach is described using this notation in Algorithm~\ref{alg:generateAssortments}. Figure~\ref{figure:collections} illustrates the generated assortments using this method for both the visual and multimodal variants.


\section{Results and Discussion}
\label{sec:results}

This section describes a large-scale experiment on Overstock~\footref{overstockurl} to determine whether simultaneously learning style from text and images provides better results than learning style from only images. All offline validations and online A/B test assume 2 variants: a visual-only assortment recommender system, which is built on top of BoVW representation of product images, and a multimodal one, which is built on text-based attributes and visual words. Our site does offer manually curated collections, but there is little overlap on the products selected in the manual process and those selected for our automated system. The methodology used to create these collections also involves selecting products from the same vendor from the same product line, which can often be viewed as substitutable products as well. These would thus serve as a poor comparison to our model, which attempts to find stylistically complimentary products of different categories regardless of vendor, rather than similar/identical products. As such we do not use them as a baseline to compare our model against.

\paragraph{Online Evaluation} We run an A/B test on Overstock~\footref{overstockurl} with both the visual variant and multimodal variant assortments on product pages. The users are provided an option to add an entire assortment or individual items from the recommendations module to their cart. User engagement with recommendations are measured via click-through rate, {\tt CTR}, on assortment recommendations is a classic measure of user engagement with product recommendations in e-commerce platforms. Our findings show that the multimodal variant outperforms the visual variant, showing a statistically significant lift of 10.9\% relative to a visual only baseline.

\paragraph{Offline Evaluations} As we are recommending sets of products together, we can't use traditional offline evaluations, such as AUC, MAP and Recall. For scoring, we have taken the assortments and scored them with the average click-based Jaccard coefficient calculated pairwise over all products contained within the assortment. We make use of logs of product clicks over the last 2 months to build this score. This tells us how compatible the various products within an assortment are to one another.

To calculate the assortment Jaccard, we look at $S$, the set of all user browsing sessions that included clicks on at least 2 items of living room furniture. A user browsing session only spans one visit, so we can assume that when items are co-clicked within the same session, the user still has the same intent, and is implicitly seeing both products as satisfying this intent. We take all user browsing sessions over a two month span for our dataset. We denote the subset of $S$ that includes sessions that have clicked on product $a$ as $S_a$.

We calculate the Jaccard coefficient for all pairs of products ($a$, $b$) within our corpus as follows:
$$J(a,b) = \frac{|S_a \cap S_b|}{|S_a \cup S_b|}$$
In contrast to the standard Jaccard formulation, if $|S_a \cup S_b| = 0$, we consider $J(a,b) = 0$. This modification is to prevent products that were never visited together from being counted as perfectly compatible with each other, since the standard formulation defaults these cases to 1.

For assortment $A$ we calculate an assortment Jaccard score $J_A$ as a simple average of the pairwise Jaccard coefficient for all items within the assortment, excluding the Jaccard coefficient between the two seeds. An example assortment with a good Jaccard score is shown in Figure ~\ref{fig:assortmentJaccard}.

\begin{table}
\label{tab:results}
  \caption{Jaccard Score averages and max values for assortments of both variants.}
    \begin{center}
\begin{tabular}{|l || c| | c| | c|}
  \hline
Modality & Avg & Max\\
\hline
\hline
visual & 0.0015 & 0.0220 \\
\hline
multimodal & 0.0027 & $\mathbf{0.0362}$\\
\hline
\end{tabular}
\label{tab:results}
\end{center}
\end{table}

The multimodal maintains a much higher average assortment Jaccard than the visual-only variant. The dramatic difference between the multimodal variant and the visual variant may reflect an underlying bias in the click data: The text modality looks at the same data used in site search and navigation to surface groups of products to users, so co-click data may reflect better discoverability of items with similar textual features as opposed to similar visual features.

We also examine the distribution of topics from each assortment. The topic representation of products in our PolyLDA generated space is more diverse than that of the LDA. The products in the LDA space usually are strongly attributed to one or two products, while the PolyLDA does a better job of representing a product as a mixture of topics. This affects style as the products within the topics that are learned are very visually similar, lending themselves to substitutability rather than complimentary behavior.

\begin{figure}[t!]
\begin{tikzpicture}[scale=0.85]
\begin{axis}[stack plots=y,ylabel=Total Assortments,xlabel=Number of Topics in Assortment, legend pos=north east]
\addplot coordinates {
(1, 2621) (2, 2343) (3, 2366) (4, 1471) (5, 1014) (6, 1073) (7, 1112) (8, 1220) (9, 1647) (10, 1787) (11, 2095) (12, 2206) (13, 2388) (14, 2438) (15, 2660) (16, 2523) (17, 1966) (18, 1294) (19, 652) (20, 278) (21, 102) (22, 21) (23, 8) (24, 1)
};
\addplot coordinates {
(1, 4245) (2, 14374) (3, 7184) (4, 2374) (5, 453) (6, 154) (7, 71) (8, 11) (9, 4) (10, 5) (11, 1) (12, 21) (13, 1)
};
\addlegendentry{Visual}
\addlegendentry{Multimodal}
\end{axis}
\end{tikzpicture}
\caption{The above chart tabulates the number of topics that non-trivially contribute to an assortment. The vast majority of visual assortments have 1-4 topics associated with them, as depicted by the peak in the red line. On the other hand, the multimodal assortments can have anywhere from one to 20 topics associated with them, which allows for more diverse product representation. This is desirable as the uncovered topics show very similar items, not necessarily complimentary ones as depicted in Figure \ref{figure:topics}. A user would likely not want an entire set of bright red furniture. Mixing among the topics to create a complimentary set is more desirable than an entire set of the exact same visual features.}
\label{figure9ugh}
\end{figure}
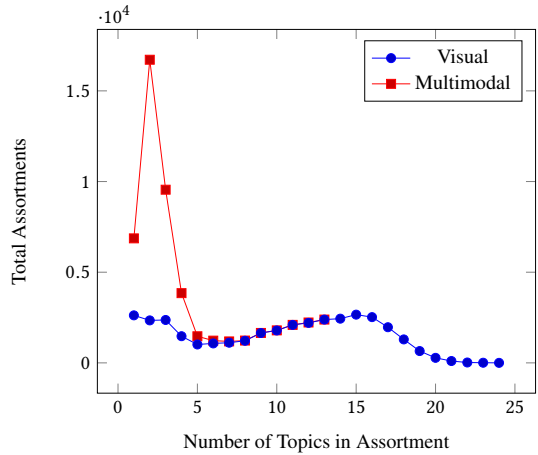

Figure~\ref{figure9ugh} shows how many topics compose each assortment. The distribution for the visual variant is heavily skewed, with the majority of assortments only being composed of a few topics. The multimodal variant has much more variety, with assortments composed of only a few topics, to assortments composed of nearly half the topics being equally likely. This shows us that the multimodal variant offers more diverse assortments. This is preferred since, as depicted in \ref{figure:topics}, the topics themselves often contain substitutable rather than complimentary items. Properly blending the topics together to create a cohesive look results in better assortments, as depicted in Figure \ref{figure:collections}.

\section{Conclusion}
\label{sec:conclusion}
Online shopping is a visual experience. Visually-aware recommendations are crucial to online shopping and e-commerce platforms. Yet, systems which rely solely on images suffer from a lack of diversity. By incorporating both image and text data we are able to create cohesive styles. 

In this paper we introduced a deep visually-aware large-scale assortments recommender system for Overstock~\footref{overstockurl}. Our assortment recommender system takes advantage of product images to create visually coherent trends from Overstock products. We introduced two variants: a visual-only variant and a multimodal variant. Our visual-only variant creates a bag-of-visual-words representation of product images by thresholding the activations from specific layers of a pre-trained deep residual neural network, {\tt Resnet-50}. It then applies topic modeling (LDA) on product image representations to create visual trends among Overstock products. We then proposed a greedy approach ( with and without budget constraints ) to create assortment recommendations based on seed items that maximize the visual compatibility of the set. Our multimodal variant takes advantage of text-based product attributes in addition to image representation. This variant utilizes Polylingual LDA (PolyLDA) to create trends that are based on two modalities, images and text. 

We have featured multiple assortments generated from both models. We have evaluated our results through a set of offline validations and an online large-scale A/B test on Overstock. Our experimental results indicate that incorporating both image and text data provides more a more cohesive visual style than using only images and can enhance user engagement metrics with recommendations module. We also show that PolyLDA provides a meaningful way to simultaneously learn style across text and image data.

\bibliographystyle{ACM-Reference-Format}
\bibliography{sigir-ecom18}

\end{document}